\documentclass[a4paper,fleqn,usenatbib]{mnras}

\usepackage[T1]{fontenc}
\usepackage{ae,aecompl}

\newcommand{\lsim}{\mathrel{\rlap{\raise -.3ex\hbox{${\scriptstyle\sim}$}}%
		  \raise .6ex\hbox{${\scriptstyle <}$}}}%
\newcommand{\gsim}{\mathrel{\rlap{\raise -.3ex\hbox{${\scriptstyle\sim}$}}%
		  \raise .6ex\hbox{${\scriptstyle >}$}}}%

\usepackage{graphicx}	
\usepackage{amsmath}	
\usepackage{amssymb}	

\title[A global map of night sky brightness]{Towards a global map of the artificial all-sky brightness }

\author[M. Kocifaj et al.]{
M. Kocifaj,$^{1}$\thanks{E-mail: kocifaj@savba.sk}
S. Bar\'{a},$^{2}$
F. Falchi$^{3,4}$
\\
$^{1}$ICA, Slovak Academy of Sciences, D\'{u}bravsk\'{a} cesta 9, 845 03 Bratislava, Slovakia\\
$^{2}$A. Astron\'{o}mica "Ío", 15005 A Coruña, Galicia\\
$^{3}$Istituto di Scienza e Tecnologia dell’Inquinamento Luminoso (ISTIL), 36016 Thiene, Italy\\
$^{4}$Departamento de F\'{\i}sica Aplicada, Universidade de Santiago de Compostela, 15782 
Santiago de Compostela, Galicia, Spain\\
}

\date{Accepted XXX. Received YYY; in original form ZZZ}

\pubyear{2015}

\begin{document}
\label{firstpage}
\pagerange{\pageref{firstpage}--\pageref{lastpage}}
\maketitle

\begin{abstract}
Modeling the hemispherical night sky brightness of anthropogenic origin is a demanding computational challenge, due to the intensive calculations required to produce all-sky maps with fine angular resolution including high-order scattering effects. We present in this Letter a physically consistent, semi-analytic two-parameter model of the all-sky radiance produced by an artificial light source that encodes efficiently the spectral radiance in all directions of the sky above the observer. The two parameters of this function are derived from the state of the atmosphere, the distance to the observer, and the source's angular and spectral emission pattern. The anthropogenic all-sky radiance at any place on Earth can be easily calculated by adding up the contributions of the surrounding artificial sources, using the information available from nighttime satellite imagery and ground-truth lighting inventories. This opens the way for the elaboration of a global world map of the artificial all-sky brightness.

\end{abstract}




\section{Introduction}
The knowledge of the hemispherical night sky brightness produced by the ever-growing expansion of outdoor lighting systems is a necessary step for characterizing the nighttime environment and monitoring the evolution of the night sky quality at present and potential astronomy observatory sites. Whereas several networks worldwide routinely gather data on the zenith night sky brightness in different photometric bands, comprehensive information on the all-sky distributions of the artificial radiance at nighttime is considerably more scarce. Excepting for a few observatories, the existing data sets are limited to the images acquired in specific observational campaigns, and the ones captured by the all-sky cameras existing in astronomic facilities for general purposes that not often include the quantitative measurement of the night sky radiance.     

In this context, modelling the artificial night sky radiance is a reasonable option. Modelling is also an instrumental tool for astronomical and environmental impact assessment of new lighting projects and territorial planning of outdoor lighting systems. The required equations are derived from radiative transfer theory and have been largely optimised for this particular field in the last two decades. Canonical examples of their outcomes are the iconic First World Atlas of the artificial night sky brightness \citep{CinzanoEtAl2001} and its updated and enhanced version \citet{FalchiEtAl2016}. These atlases display the brightness of the zenith sky in the Johnson-Cousins V band, for a world grid with a ground pixel resolution of 30 arcmin.  

Modelling the all-sky, hemispherical radiance distribution produced by artificial lights is however a more demanding challenge. Generating a fine angular resolution sky map requires computing the artificial brightness in a large number of directions on the sky. This is a computationally expensive task, especially if narrow band spectral resolution is required and higher orders of scattering are to be included. One way of reducing the dimensionality of the problem is expressing the hemispheric radiance as a linear combination of suitable basis functions or modes as, e.g. Zernike or Legendre polynomials \citep{BaraEtAl2015a,BaraEtAl2015b}. This approach allows to compress the information of a ~10$^{6}$ pixel map into ~10${^2}$ modal coefficients, enabling a significant reduction of computational load. However, these polynomial bases were not specifically designed for describing the typical patterns of the artificial night sky, and hence are not expected to be the optimal ones for this task.

In this work we develop a two-parameter model that efficiently matches the spectral radiance in all directions of the sky above the observer calculated by means of accurate numerical procedures. This model allows for an additional reduction of two orders of magnitude in the number of required parameters, in comparison with previous attempts based on polynomial bases. The model is physically grounded, not merely heuristic, and its two parameters are related by means of analytic expressions or look-up tables to the state of the atmosphere, the distance to the observer, and the source's angular and spectral emission pattern. The analytic expression for the sky radiance is formally coincident with the one described in \citet{KociBara2019}; however, the two parameters $t$ and $g$ are now generalized to encompass higher-order scattering effects (up to the 5th order), and are provided ab initio, as a part of the model, not estimated a posteriori from images captured on site.

\section{The model}
\label{sec:model}
Unlike empirical approaches, we model the all-sky radiance as a function of two parameters, $g$ 
and $t$ which are commonly used to characterize the atmospheric environment \citep{KociBara2019}. 
In radiative transfer theories $g$ has relation to the mean asymmetry parameter of a turbid 
atmosphere, while the physics interpretation of $t$ is possible via mean optical attenuation 
along the beam path from a light source to a measuring site. For a cloudless optically thin atmosphere 
$g$ is derived from aerosol ($P_{a}$) and Rayleigh ($P_{R}$) scattering phase functions using 
aerosol ($k_{a}$) and Rayleigh ($k_{R}$) volume extinction coefficients as weighting factors 
(see e.g. Eq. 5 in \citep{KatkovskyEtAl2018})
\begin{eqnarray}\label{eq:weighted_P}
    P(g,\theta) &=& \frac{P_{a}(g_{a},\theta) \omega_{a} k_{a}+P_{R}(\theta)k_{R}}{k_{a}+k_{R}}~~~.
\end{eqnarray}
Here $\omega_{a}$ is the aerosol single scattering albedo and $\theta$ is the scattering angle -- 
i.e. angle contained by the directions of incident and scattered waves. Alike in the Earth's 
atmosphere, the above approach is also applied to various complex discrete media, including 
particles dispersed in e.g. seawater (e.g. Eq. 78 in \citet{FokouEtAl2021}). The formula 
\ref{eq:weighted_P} has a solid theoretical foundation in studying light scattering processes 
at a low scale, involving small atmospheric volumes. However, multiple scattering of light along 
with non-uniform beam attenuation at inclined trajectories introduce multifaceted distortions 
to NSB patterns which are difficult to describe using Eq.~\ref{eq:weighted_P}. Therefore, the 
total scattering phase function, $P(g,\theta)$, in a clear atmosphere (for $k_{a}=0$) is highly 
unlikely to mimic that of $P_{R}(\theta)$. Although $P_{R}(\theta)$ shows ideal symmetry for 
forward and backward hemispheres with the peak ratio $P_{R}(0^{\circ})/P_{R}(90^{\circ})$=2, 
the radiance in multiply scattering Rayleigh atmosphere spans a wider range, resulting in  
$P_{R}(0^{\circ})/P_{R}(90^{\circ})$>2 (see also leftmost plot in Fig. 12 in \citet{Mobley2015} 
for sky elements at a horizontal circle; i.e. for zenith angle $z$ being fixed). Assuming that 
$z$ is conserved, the changes to the sky radiance $L(z,A)$ are driven only by the scattering angle 
$\theta$, as it follows from the model by \citet{KociBara2019} 
\begin{eqnarray}\label{eq:radiance}
    L(z,A) &=& L_{S} P(g,\theta) \frac{(1-g)^2}{(1+g)} \frac{M(z)}{M_{S}t} \frac{e^{[M_{S}-M(z)]t}-1}{M_{S}-M(z)}~~~.
\end{eqnarray}
Looking at the horizon, the above formula reduces to $L_{S} P(g,\theta) \frac{(1-g)^2}{(1+g)}$. In Eq.~\ref{eq:radiance} $z$ and $A$ are the observational zenith and azimuth angles, respectively. $L_{S}$ is the radiance leaving the source in the azimuth of the observer. We assume azimuthal symmetry of the source emissions.
Analogously to the solution concept we have implemented in \citet{KociBara2019}, we use the Henyey-Greenstein (HG) function to describe the shape of the total scattering 
phase function $P(g,\theta)$. The HG function is also known to provide reasonably accurate predictions 
for aerosol ensembles, provided that the asymmetry parameter $g_{a}$ used is accurate 
\citep{KahnertEtAl2005}. Therefore we substitute $P_{a}(g_{a},\theta)$ for its HG-equivalent. 
The Rayleigh scattering phase function $P_{R}(\theta)$ can be expressed analytically in the form 
of $\frac{3}{4}(1+\cos^{2}\theta)$. The functions $M(z)$ and $M_{S}$ in Eq.~\ref{eq:radiance} 
are the optical air mass in directions of $z$ and the light source, respectively. 

It is reasonable to expect that in a non-turbid, but still multiple-scattering atmosphere, $P(g,\theta)$ still displays some 
degree of anisotropy which can be modelled by $P_{0}(g_{0},\theta)$ with $g_{0}>0$. We assume 
that in a turbid atmosphere $P(g,\theta)$ be a non-trivial superposition of the basis functions 
$P_{0}(g_{0},\theta)$ and $P_{a}(g_{a},\theta)$.

\section{Inputs to the model}
\label{sec:inputs} 
An exceptional simplification of otherwise vast numerical modelling of NSB distributions arises 
from the separation of variables concept, which is a peculiar property of the model developed. 
The key element which makes the model really strong is the proper adjustment of the input parameters 
$t$ and $g$. \\
~\\
{\it The parameter $t$}\\
Due to its nature  the product of $t$ and $M_{S}$ determines the optical transmission 
coefficient ($e^{M_{S}t}$) of the atmospheric volume between the source of light and the observer. 
Assuming the beam of light traverses the atmosphere horizontally, the intensity decays 
proportionally to $e^{-T}$, where $T=(k_{a}+k_{R})D$; $D$ being the separation distance 
between source of light and observer. By satisfying an identity criterion for both exponential 
functions, we obtain 
\begin{eqnarray}\label{eq:parameter_t}
    t &=& \left ( \frac{\tau_{a}}{H_{a}}+\frac{\tau_{R}}{H_{R}} \right ) \frac{D}{M_{S}}~~~,
\end{eqnarray}
where $\tau_{a}$ and $H_{a}$ are the aerosol optical thickness of a vertical atmospheric column 
and the aerosol scale height, respectively. The respective parameters for air molecules are 
$\tau_{R}$ and $H_{R}$. Eq.~\ref{eq:parameter_t} is valid for an exponential atmosphere with 
vertical stratification of aerosol concentration being proportional to $k_{a} e^{-h/H_{a}}$; 
$h$ is the altitude above the ground. The vertical distribution of air molecules is modelled 
analogously, so $k_{R} e^{-h/H_{R}}$.\\
~\\
{\it The parameter $g$}\\
No suitable analytical expression for parameter $g$ exists yet. We have retrieved $g$ by
matching the modelled NSB distributions to the ones obtained from highly accurate multiple 
scattering computations \citep{Kocifaj2018}. The latter allows for modelling the higher-scattering 
diffuse radiance of the night sky at arbitrary altitude, horizontal separation, and spectral band. 
It is highly important to determine $g$ in a narrow spectral band, first because $g$ is wavelength-dependent, and also because the exact solution to the governing equations (and underlying 
Maxwell equations) normally requires the concept of perfectly monochromatic radiation. 

Here we show the solution for wavelength $\lambda$=550 nm, located in the middle of the visible 
spectrum. Considering a relatively smooth variation of the atmospheric scattering properties, the 
results we obtain are also representative for adjacent wavelengths, roughly located in a  
spectral band of width $\pm$20-30 nm. An advantageous feature of such computations is that 
the above spectral band overlaps with the dominant emissions from LPS, HPS, MH, or MV
lights, and also with the phosphor re-emission peak of pc-LED spectra.

\begin{figure}
	\includegraphics[width=0.8\columnwidth]{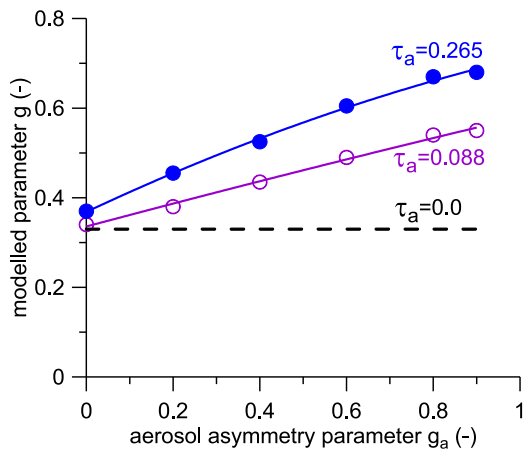}
   \caption{The parameter $g$ as a function of aerosol asymmetry parameter $g_{a}$ for three 
   discrete values of aerosol optical depths ($\tau_{a}$).}
   \label{fig:parameter_g}
\end{figure}

The optimum value of $g$ is determined by minimizing the differences between the radiance distribution 
modelled from Eq.~\ref{eq:radiance} and the one computed as a sum of five scattering orders \citep{Kocifaj2018}. 
The parameter $g$ is shown in Fig.~\ref{fig:parameter_g} as a function of $g_{a}$ for three discrete
aerosol optical depths. The results are in conformity with our premise that $P(g,\theta)$ asymptotically 
approaches that of $P_{0}(g_{0},\theta)$ with $g_{0}\approx0.33>0$; which excludes isotropic scattering 
patterns even if the aerosol optical thickness is nearly zero. The approximate formula we have found for 
$g$ is as follows
\begin{eqnarray}\label{eq:approximate_g}
    g &=& c_{0}+c_{1}g_{a}+c_{2}g_{a}^{2}~~~, 
\end{eqnarray}
with 
\begin{eqnarray}\label{eq:scaling_c}
    c_{0} &=& 0.33+0.15 \tau_{a} \nonumber \\
	c_{1} &=& 0.9 \tau_{a}^{0.51} \nonumber \\
	c_{2} &=& 1.3 \tau_{a}^{1.85}~~~.
\end{eqnarray}

~\\
{\it Input data}\\
Ground-based monitoring networks such as the Aerosol Robotic Network (AERONET) normally provide the best
opportunities to characterize aerosols in the lower atmosphere. Unfortunately, ground stations are 
sparsely distributed, making it difficult a global aerosol mapping \citep{WeiEtAl2020}. However, 
aerosol satellite products with  wide-scale coverage are a useful source of information on global
distribution of $\tau_{a}$ and $g_{a}$ \citep{Shikwambana2018}. Satellites orbiting the Earth collect 
real-time data, thus updating aerosol parameters dynamically by region and season \citep{RemerEtAl2005}. 
Consistent records of the Earth's aerosol system are available through the Moderate Resolution Imaging 
Spectroradiometer (MODIS) on NASA's Terra satellite \citep{Modis2022,Terra2022}. Missing aerosol data 
can be interpolated or inferred based on the aerosol types persistently prevalent in the region/season 
or based on the information on dominant aerosol emission sources in a local or regional context. This is possible
since a global aerosol climatology project has shown that the aerosol optical properties at 550 nm 
depend markedly on aerosol types, specifically on whether fine-mode or coarse-mode particles dominate
the regional emissions. The project has also analysed aerosols of anthropogenic origin, dust, sulfates, 
nitrates, sea-salts and organic-matter \citep{KinneEtAl2013,Kinne2019}.

~\\
{\it Artificial light sources}\\
The radiance of the light sources surrounding the observer is another key input for calculating the 
artificial all sky brightness. Monochrome and RGB imagery of artificial light emissions with nearly 
worldwide coverage is presently provided by several on-orbit radiometers  \citep{ElvidgeEtAl2017,ElvidgeEtAl2021,LevinEtAl2020,LiEtAl2022,ZhengEtAl2018} as well as by the Crew 
Earth Observation program of the International Space Station \citep{SdM2019,SdM2021,StefanovEtAl2017}. 
These inputs are expected to be enough for computing world hemispheric sky brightness maps with a level 
of accuracy comparable to that already attained by their zenith counterparts \citep{FalchiEtAl2016}. 
Earth observation at nighttime is a thriving field and it can be anticipated that observing platforms 
with enhanced spectral and angular sensing capabilities will be planned and developed in the next years.

The final calculation of the hemispheric radiance, for any given observing site, is made by adding up 
the individual hemispheric contributions of each light source. Within the range of validity of the model 
here presented, the contributions of sources located at equal distances from the observer but different 
azimuths are just rotated versions of the same basic pattern, weighted by the radiance emitted by each 
source in the direction to the observer. The computational burden can be further alleviated by 
pre-calculating the model parameters for a dense set of different distances, atmospheric conditions, 
angular and spectral source emission patterns and observation photometric bands.
   
\section{Model corroboration}
\label{sec:validation} 
We have used the functionality of the multiple scattering code MSOS1 \cite{Kocifaj2018} to accurately model 
the night sky radiance distributions for the wavelengths of 550 nm and 450 nm, two aerosol scale
heights ($H_{a}$=1.5 km and 2.2 km) and a set of discrete aerosol asymmetry parameters ($g_{a}$) ranging 
from 0 to 0.9. Single scattering albedo of aerosols $\omega_{a}$ is typically as large as 0.95 which 
allows for substitution of volume scattering coefficient $\omega_{a} k_{a}$ for volume extinction 
coefficient $k_{a}$ in our numerical runs and thus reduce the degrees of freedom. The exact computations
are performed up to fifth scattering order to guarantee that multiple scattering radiances are accurate 
to within a few tenths of percent for all grid points on the modelling domain. 

\begin{figure}
	\begin{minipage}[t]{0.49\columnwidth}
	\includegraphics[width=1\columnwidth]{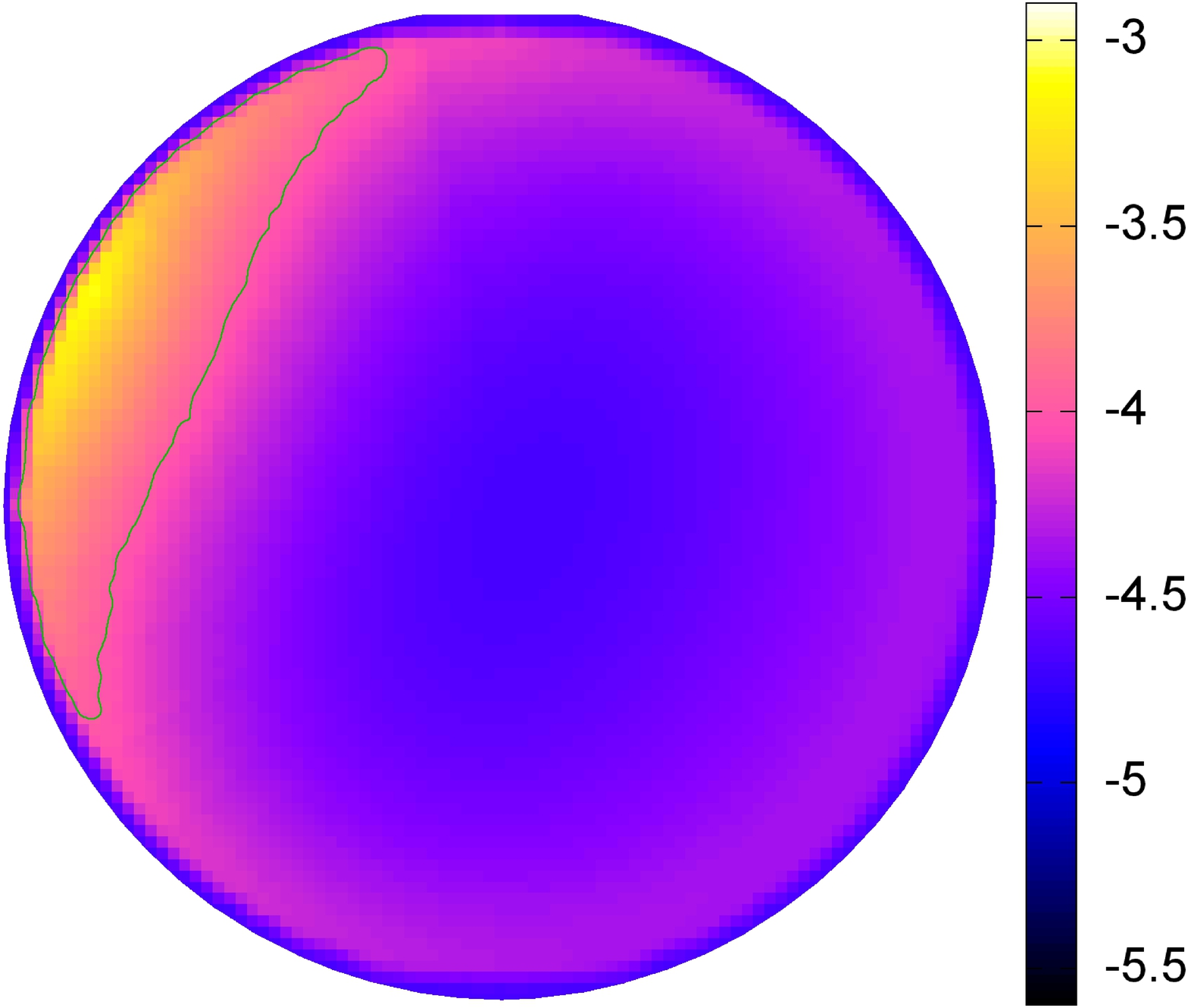}
	\end{minipage}
	\begin{minipage}[t]{0.49\columnwidth} 
	\includegraphics[width=1\columnwidth]{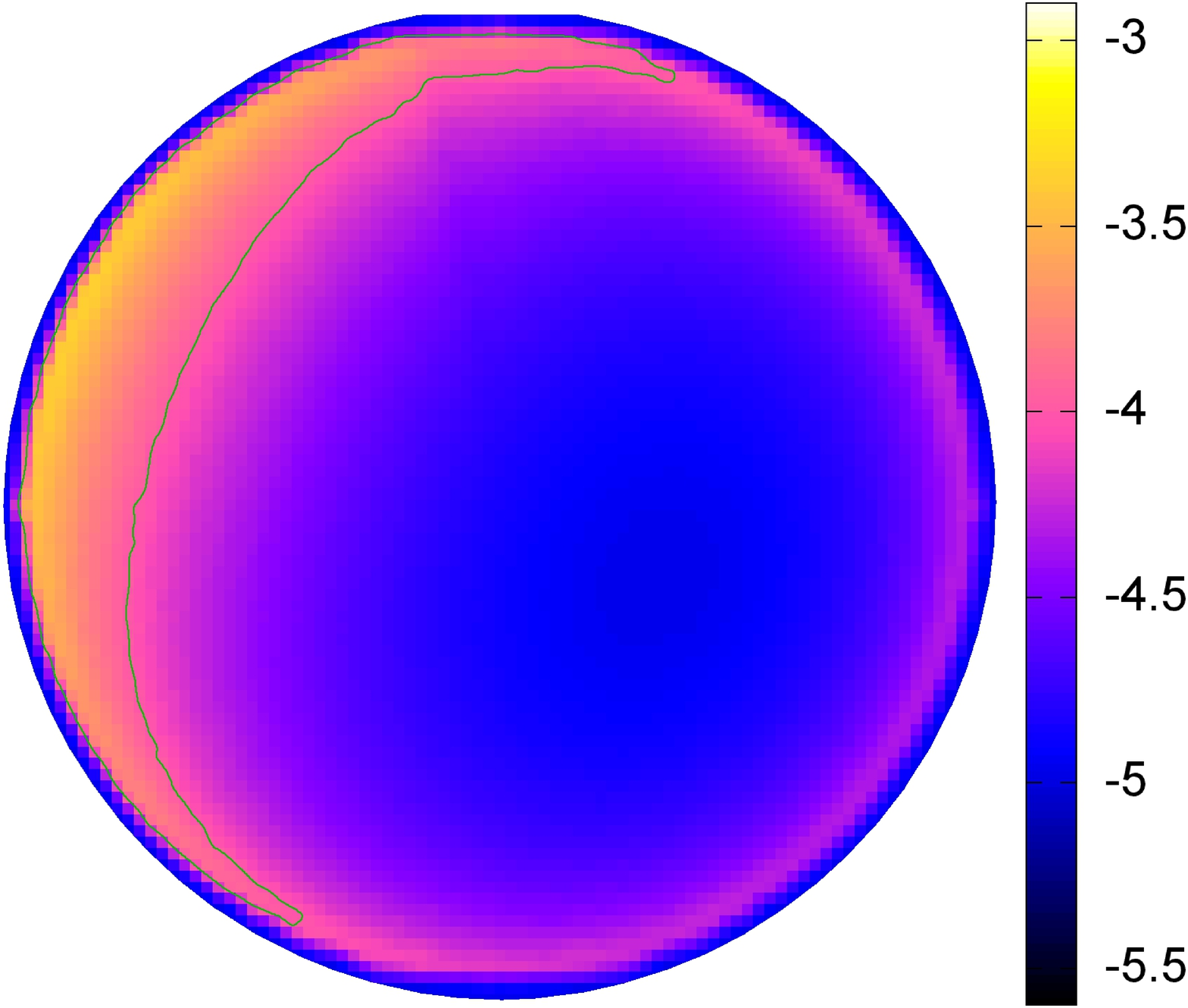}
	\end{minipage} 
	\begin{minipage}[t]{0.49\columnwidth}
	\includegraphics[width=1\columnwidth]{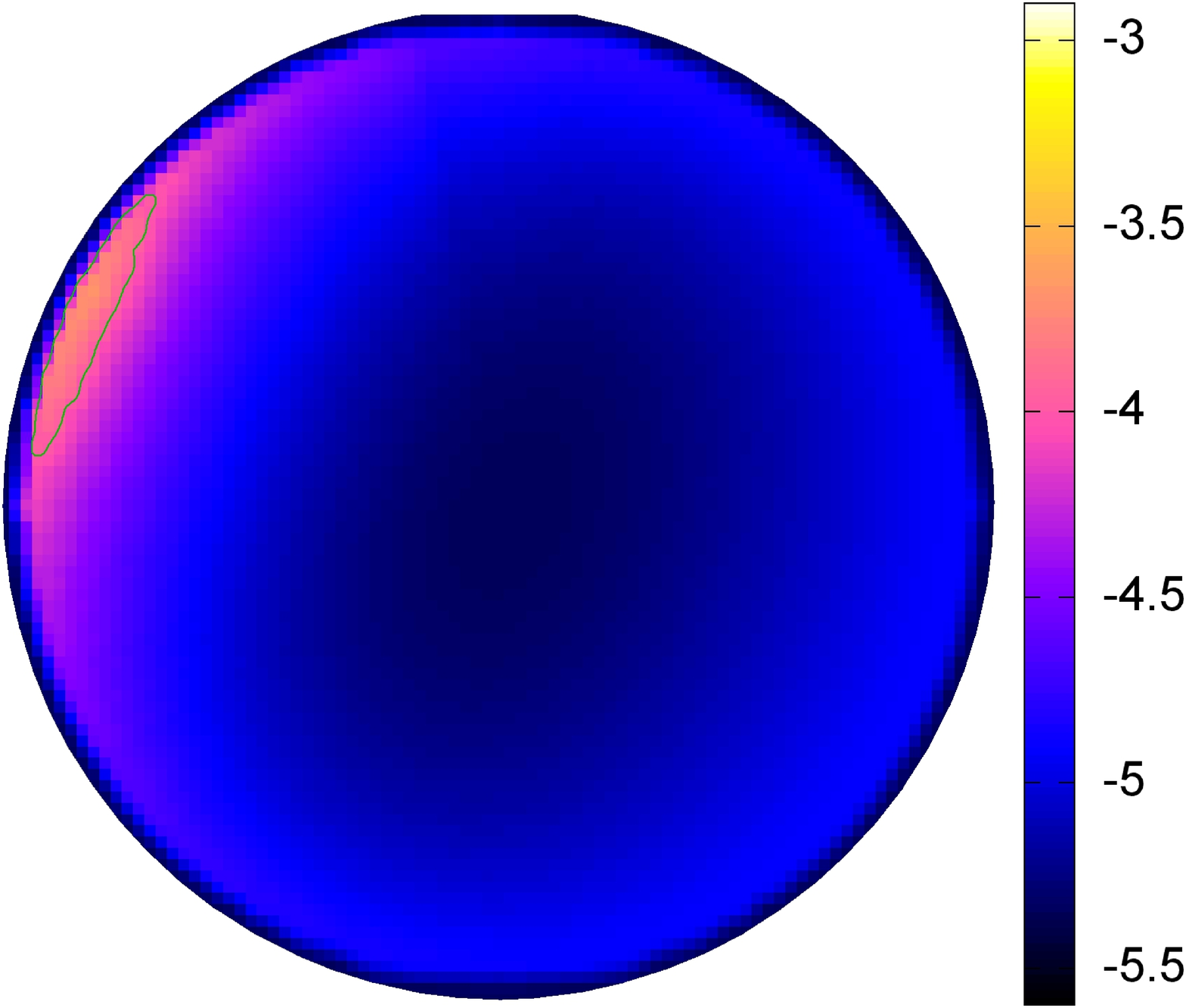}
	\end{minipage}
	\begin{minipage}[t]{0.49\columnwidth} 
	\includegraphics[width=1\columnwidth]{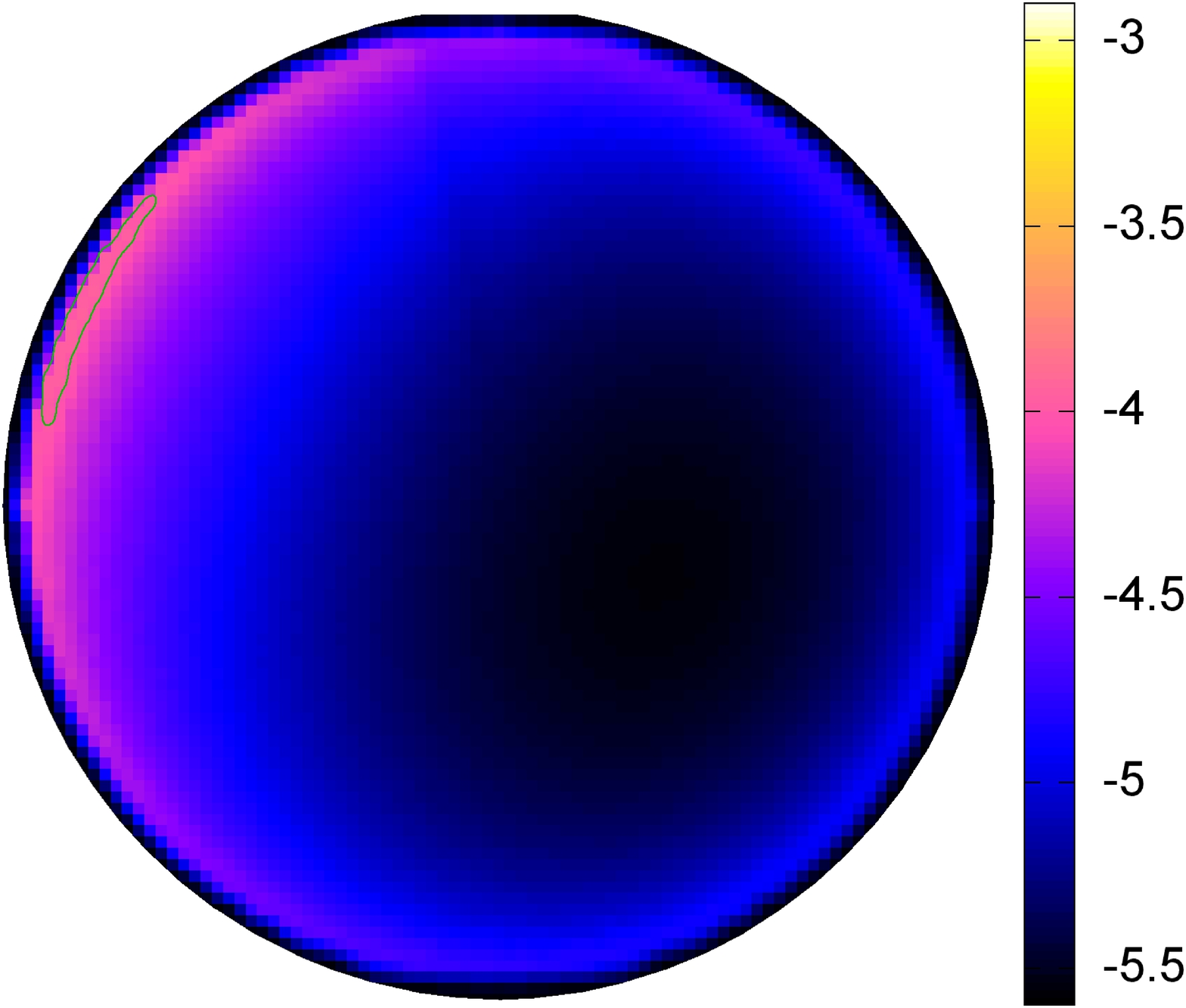}
	\end{minipage} 
   \caption{All plots are false color images of the night sky radiance distributions at 450 nm for a light 
   source seen at the azimuth angle of 294$^{\circ}$. For each figure the azimuth is measured in a clockwise 
   direction (with north at the top). The zenith and horizon ($z=85^{\circ}$) are in the center and at the 
   edge of each plot, respectively. Left column shows results of exact computations using a multiple scattering 
   code \citep{Kocifaj2018}, right column is for the model (Eq.~\ref{eq:radiance}). Top panels are for the 
   horizontal distance of $D$=7km; bottom panels are for $D$=15km. Radiance data in arbitrary units are plotted 
   on a logarithmic scale. The aerosol optical depth is $\tau_{a}$=0, while the remaining input data are in described in the 
   main text.}
   \label{fig:AOD0_450nm}
\end{figure}

We first validate the model for an aerosol-free atmosphere ($\tau_{a}=0$). In this case the results 
should be independent of any aerosol properties. This not only makes the model validation in limit 
conditions possible, but also provides an ideal starting point for increasingly detailed study. 
Fig.~\ref{fig:AOD0_450nm} documents that the modelled radiance distribution at 450 nm (right column) 
matches the exact computations 
(left column). The radiance data are shown on the same 
logarithmic scale to allow for reasonable comparison. The best fit parameter $g_{0}=0.36$ for the 
wavelength of 450 nm is a bit higher than that for 550 nm (see Fig.~\ref{fig:parameter_g}). However, 
this finding is fully consistent with what has been indicated earlier -- specifically that $g$ (and 
$g_{0}$ as well) should exhibit spectral features. The mean discrepancy of modelled and computed 
radiance averaged over all sky elements is 20\%. 

\begin{figure}
	\begin{minipage}[t]{0.49\columnwidth}
	\includegraphics[width=1\columnwidth]{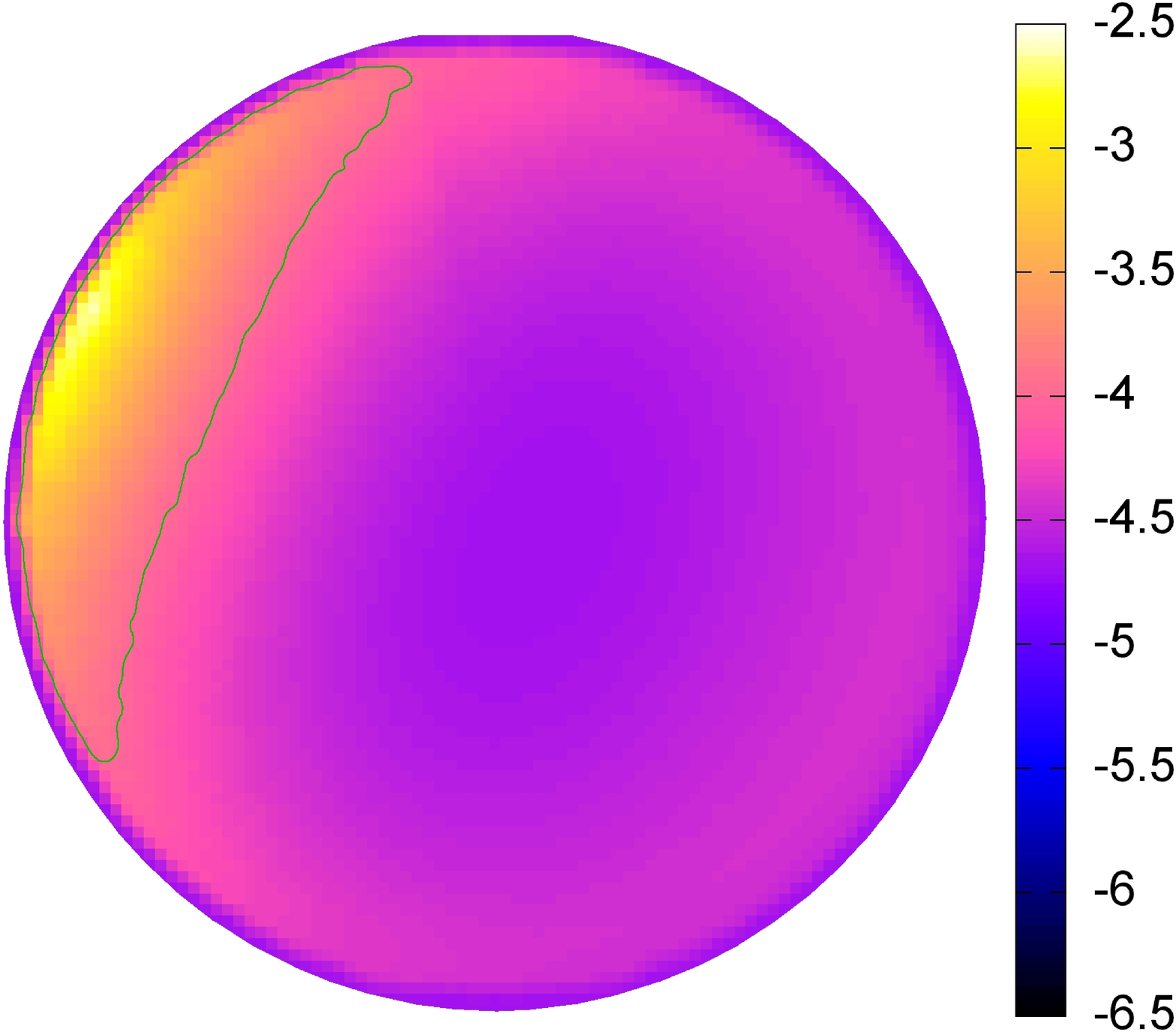}
	\end{minipage}
	\begin{minipage}[t]{0.49\columnwidth} 
	\includegraphics[width=1\columnwidth]{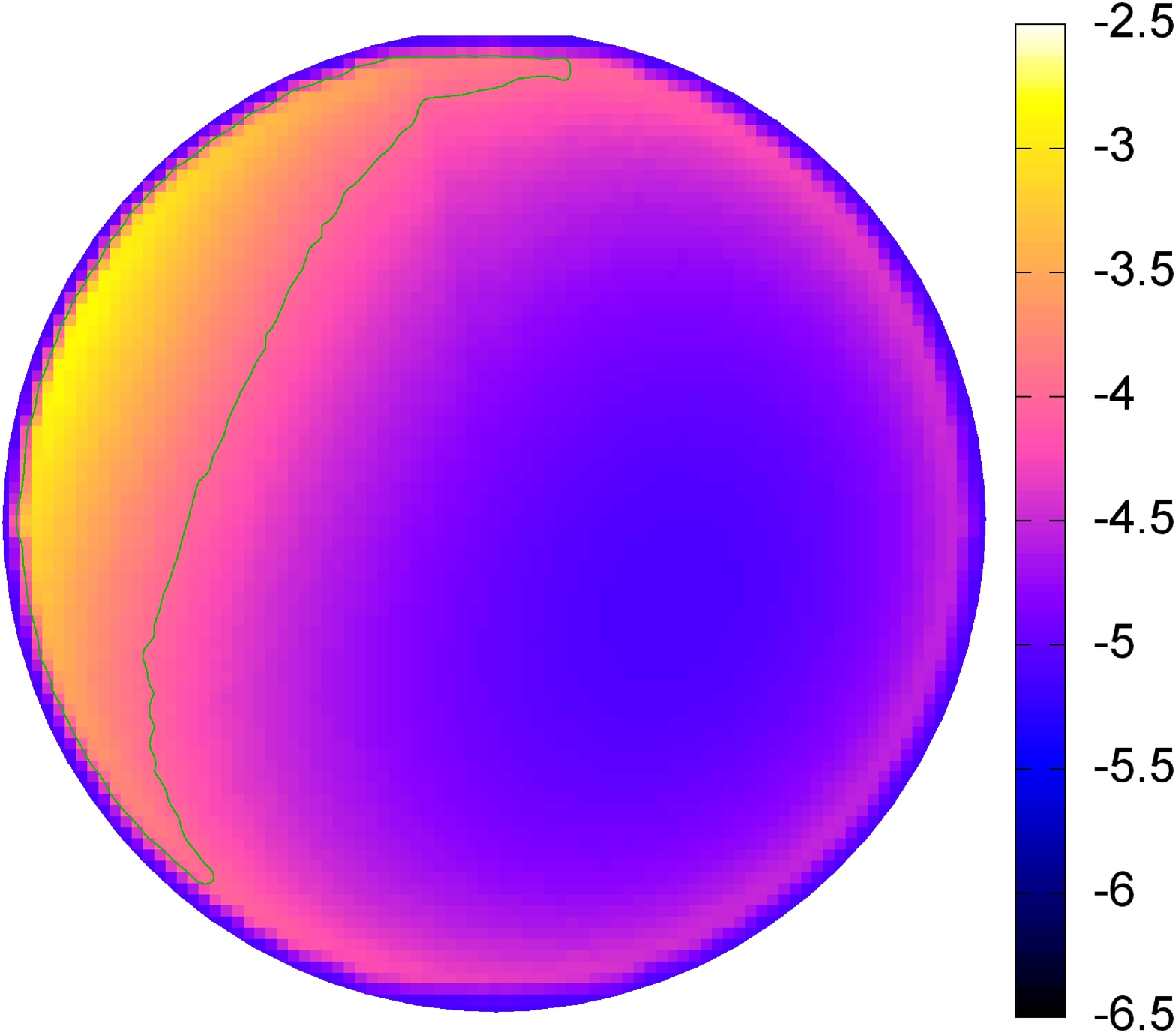}
	\end{minipage} 
	\begin{minipage}[t]{0.49\columnwidth}
	\includegraphics[width=1\columnwidth]{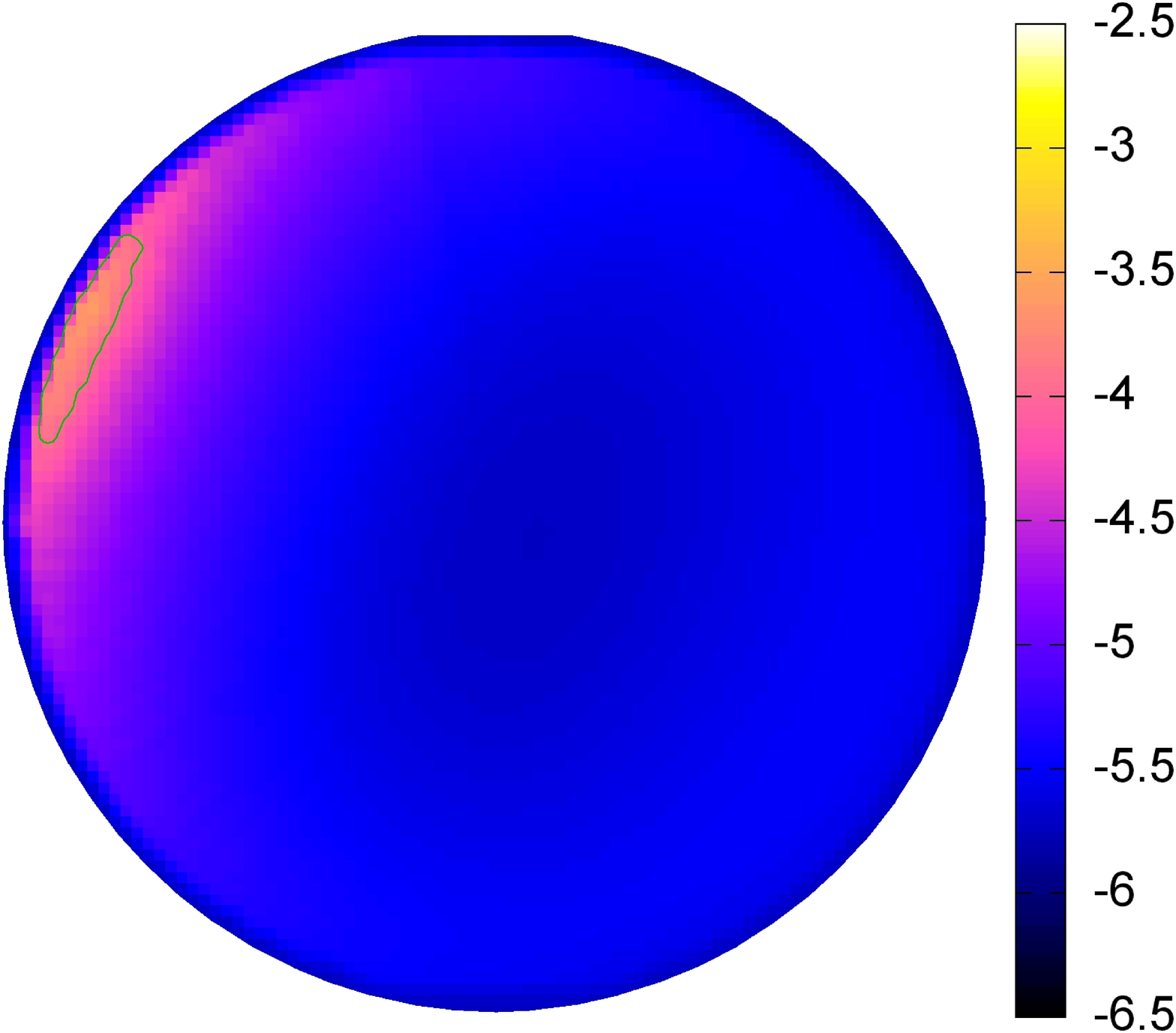}
	\end{minipage}
	\begin{minipage}[t]{0.49\columnwidth} 
	\includegraphics[width=1\columnwidth]{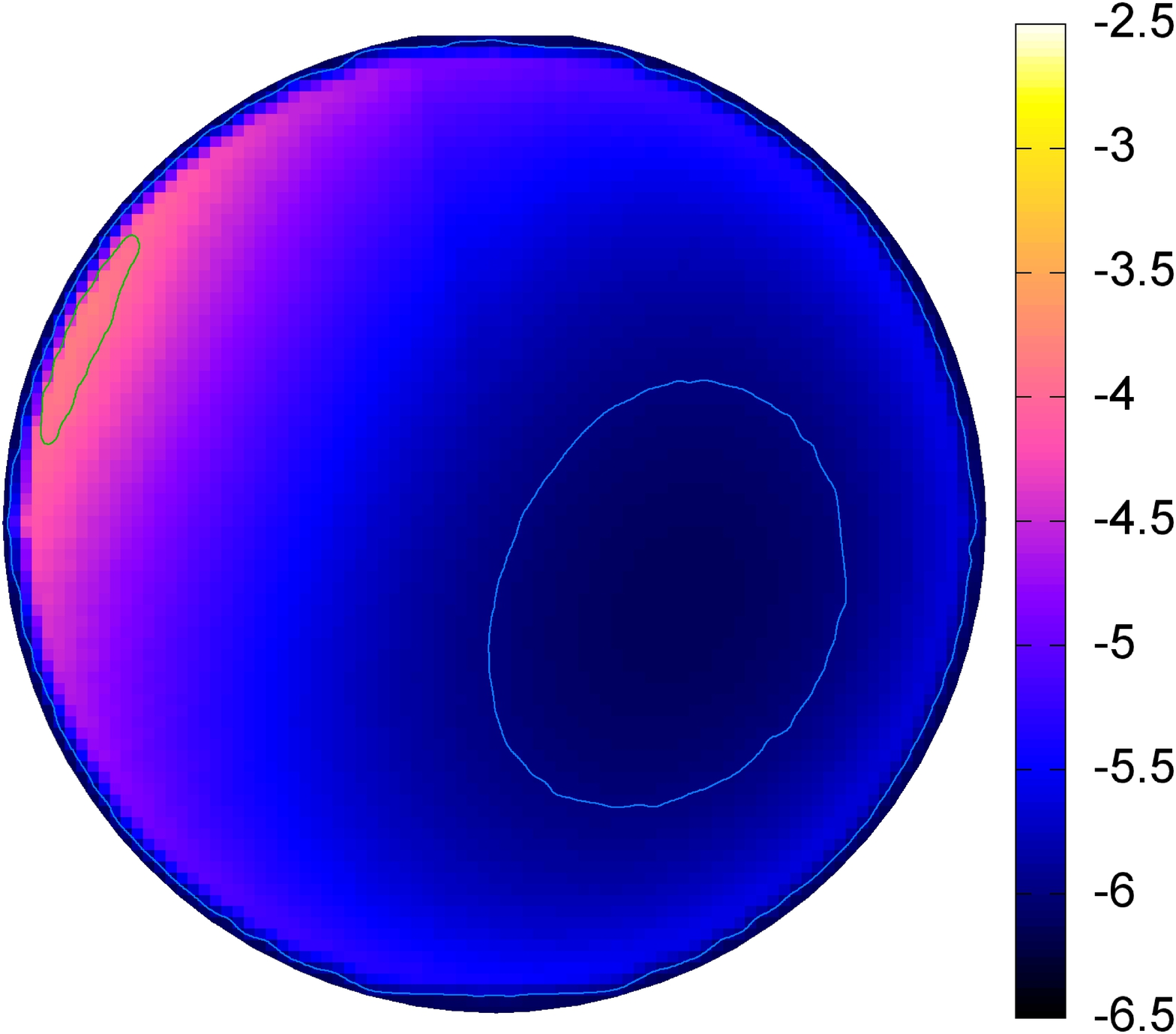}
	\end{minipage} 
   \caption{The same as in Fig.~\ref{fig:AOD0_450nm}, but for $\tau_{a}$=0.23 and aerosol asymmetry 
   parameter $g_{a}$=0.85.}
   \label{fig:AOD023_450nm}
\end{figure}

The same type analysis is being applied to $\tau_{a}=0.23$, which corresponds to a moderately 
polluted atmosphere. AODs around 0.2 not only belong to the most abundant class of aerosol 
optical depths (see e.g. \cite{MarkowiczEtAl2021}), but are also representative for marine and 
continental aerosol systems \citep{WeltonEtAl2002}. Fig.~\ref{fig:AOD023_450nm} demonstrates the 
model's ability to produce reasonable predictions for a turbid atmosphere, with even a highly 
anisotropic scattering function $P(\theta)$ and radiance amplitudes varying over several 
orders of magnitude. It is well documented that the aerosol particles make $P(\theta)$ strongly 
forward-lobed, which along with increased extinction at the blue edge of the visible spectrum results 
in transitioning from slight to steep NSB gradation. To demonstrate the effect of increased 
forward scattering from large particles we conducted numerical experiment on $P_{a}(g_{a},\theta)$ 
with $g_{a}$=0.85, which is a typical vaue for water aerosols \citep{PengLi2016,GraafEtAl2005}. 
Large particles may have tendency to concentrate in the lower atmosphere, thus we chose $H_{a}$ as 
low as 1.5 km. The mean discrepancy between modelled and accurately computed radiance distributions 
is 20-25\%. The solution obtained for heavily polluted atmosphere ($\tau_{a}$=0.57, not shown here) 
keeps the same error margin. 

\begin{figure}
	\begin{minipage}[t]{0.49\columnwidth}
	\includegraphics[width=1\columnwidth]{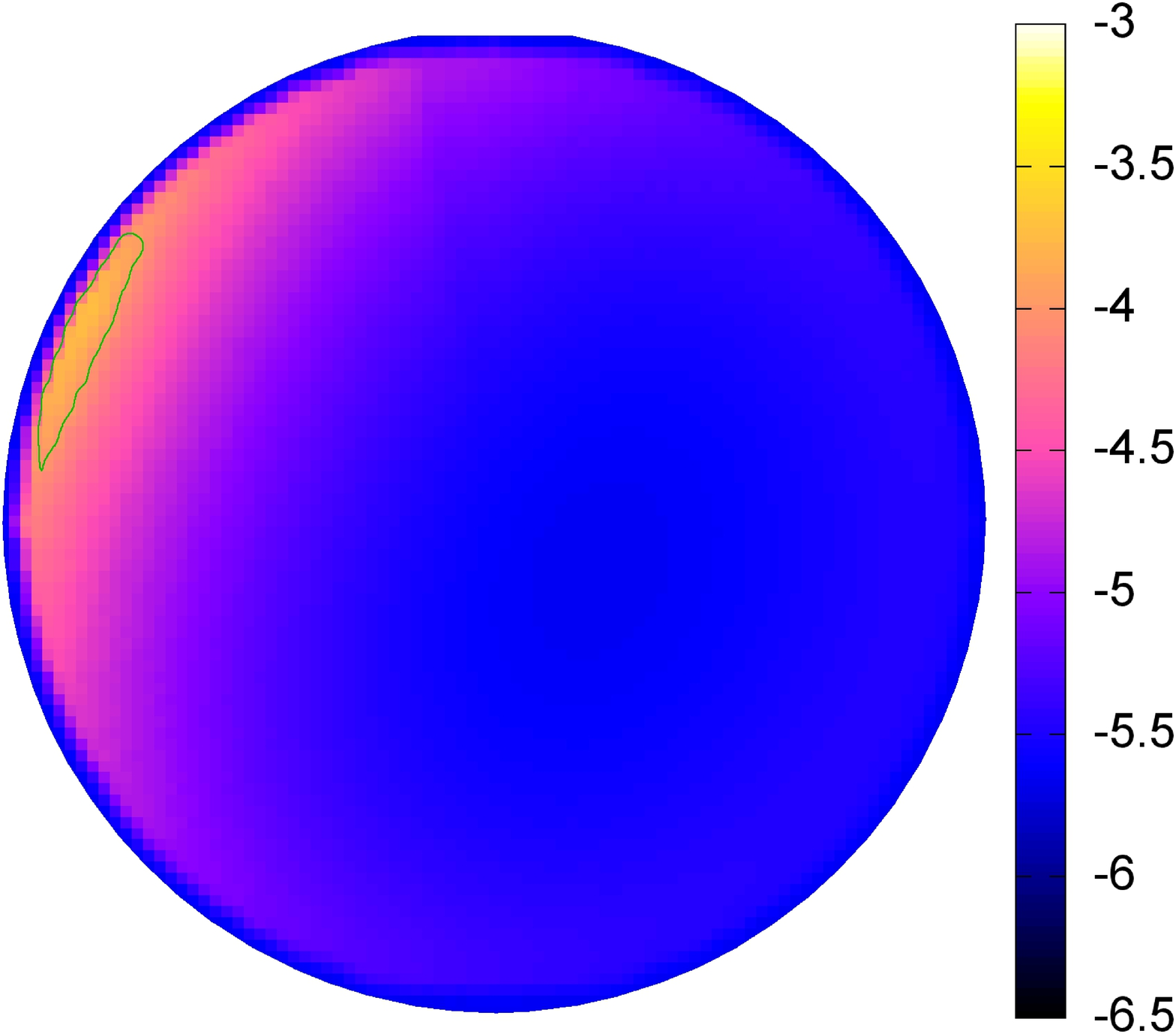}
	\end{minipage}
	\begin{minipage}[t]{0.49\columnwidth} 
	\includegraphics[width=1\columnwidth]{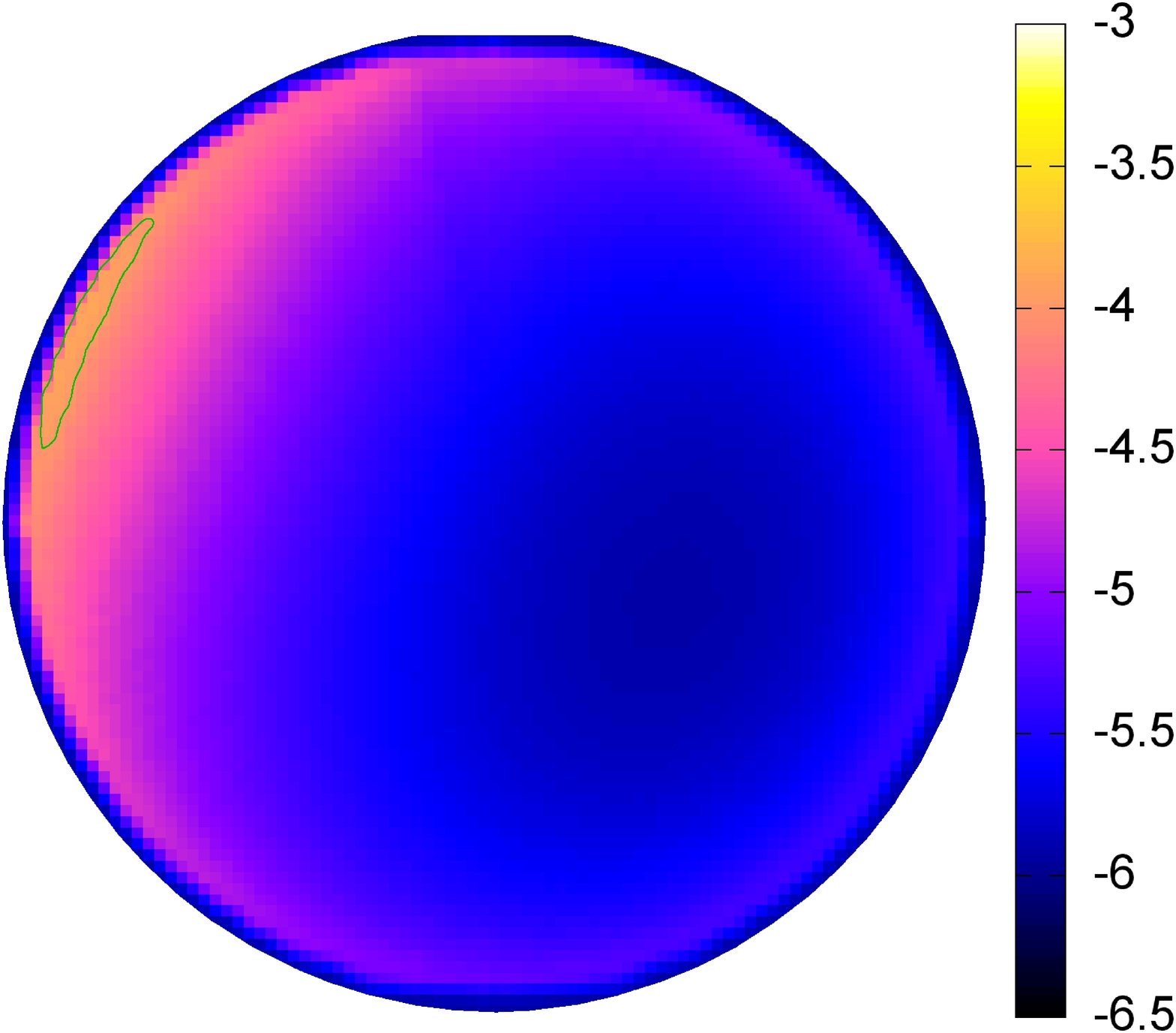}
	\end{minipage} 
	\begin{minipage}[t]{0.49\columnwidth}
	\includegraphics[width=1\columnwidth]{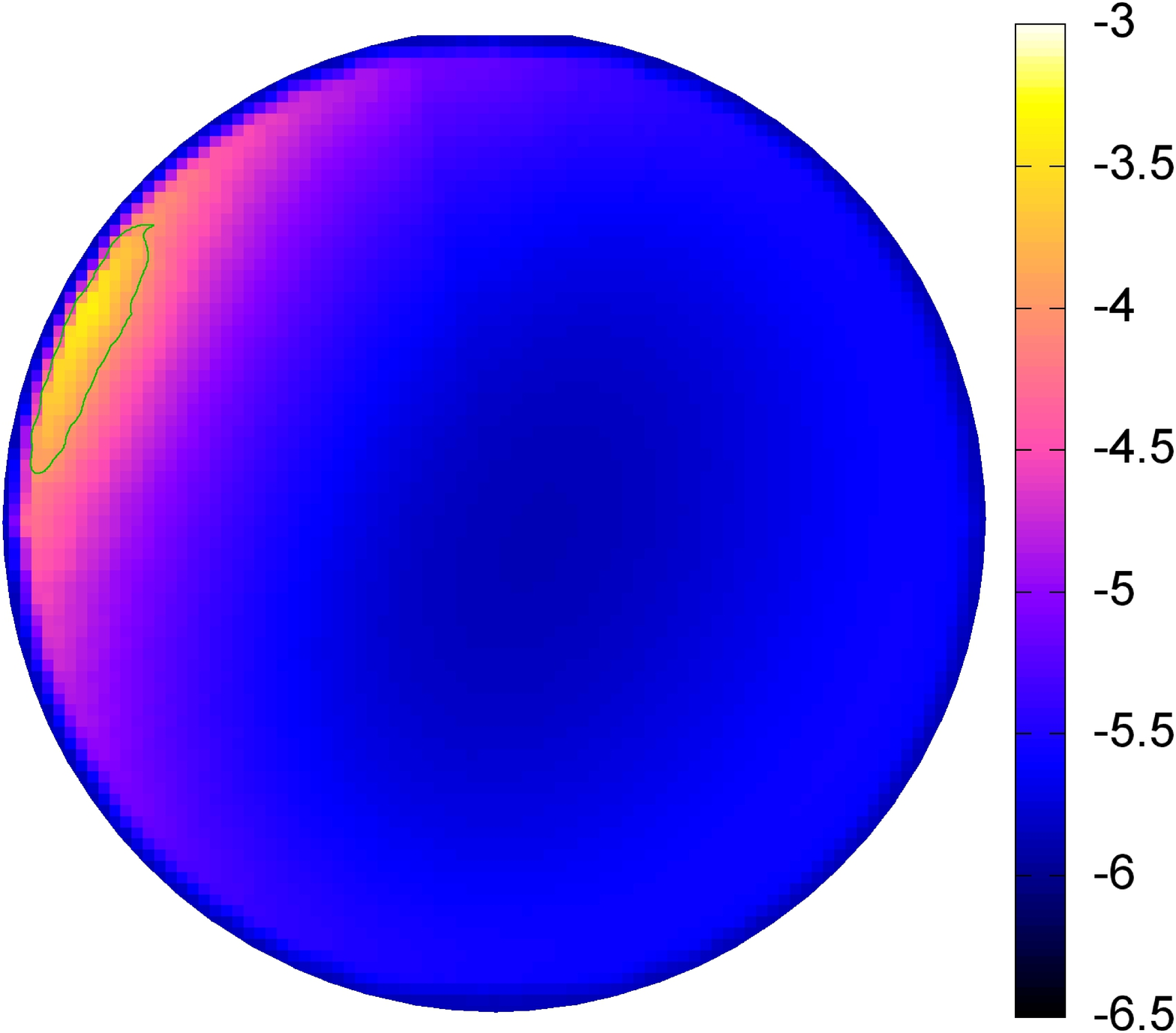}
	\end{minipage}
	\begin{minipage}[t]{0.49\columnwidth} 
	\includegraphics[width=1\columnwidth]{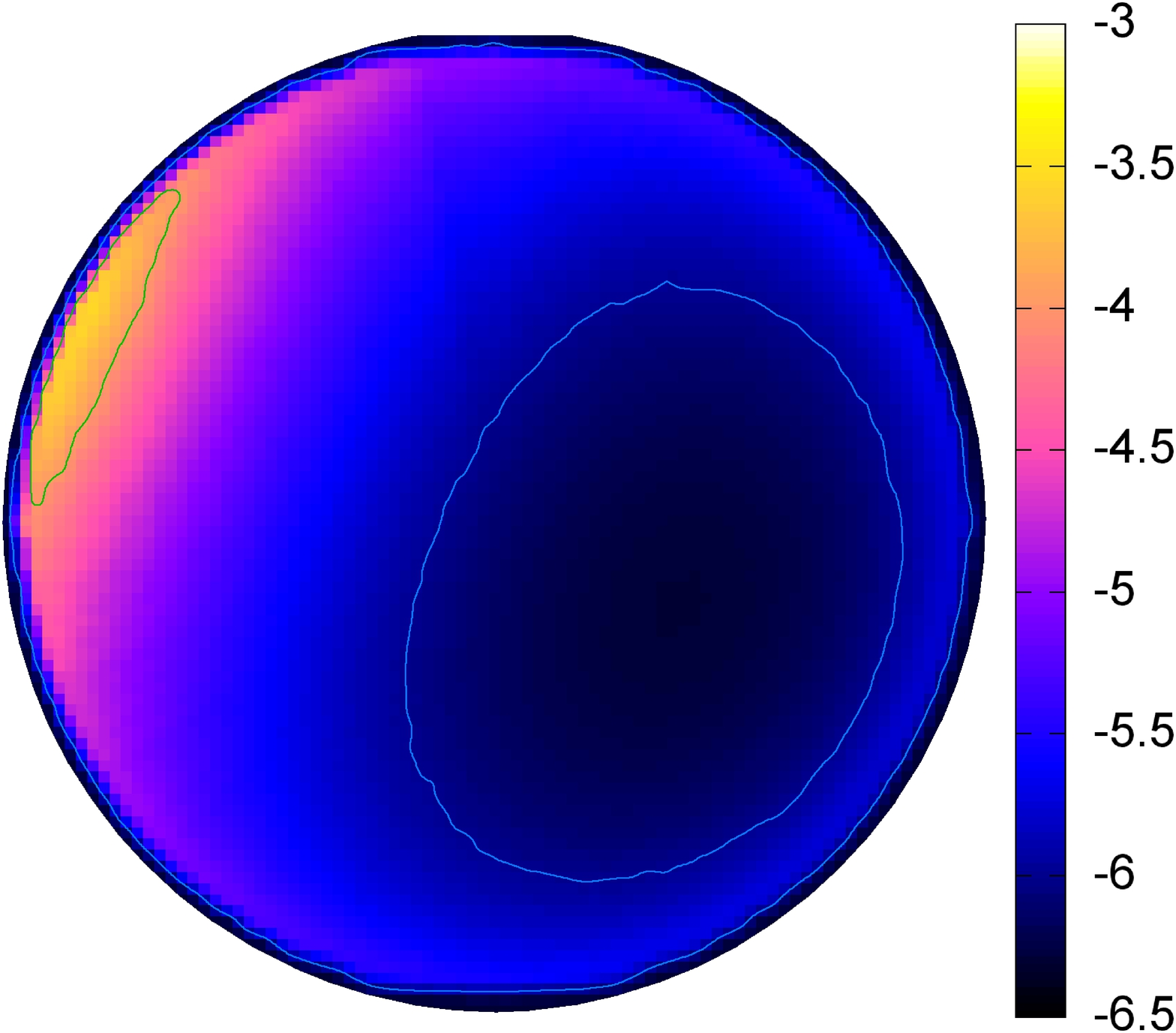}
	\end{minipage} 
   \caption{The same as in Fig.~\ref{fig:AOD0_450nm}, but for wavelength of 550 nm, $\tau_{a}$=0.265, 
   $H_{a}$=2.2 km, and $D$=15km. Top panels are for aerosol asymmetry parameter $g_{a}$=0.4, while the
   bottom panels are for $g_{a}$=0.9.}
   \label{fig:AOD0265_650nm}
\end{figure}

Systematic computations for green light (550 nm) using multiple scattering tool include situations 
with $\tau_{a}$=(0.0, 0.088, 0.265),  $g_{a}$=(0, 0.2, 0.4, 0.6, 0.8, 0.9), and $D$=(1.3, 2.9, 6.7, 
15.4, 35.3, 81.3). The overall deviation of modelled from accurately computed NSB distributions 
ranges from 15\% to 25\%. The proposed model was successful to simulate low and high aerosol contents,
even the aerosol systems with $\tau_{a}$ exceeding $\tau_{R}$ by a factor of 2-3. The model works
well for all aerosol asymmetry parameters studied, as documented in Fig.~\ref{fig:AOD0265_650nm} 
for $g_{a}$=0.4 and 0.9. This make us confident that the model can significantly compensate the 
effort that otherwise would be needed for systematic long-term sky surveys worldwide.

\section{Discussion}
\label{sec:discussion} 

The analytical expression in  (Eq.~\ref{eq:radiance}), with the appropriate values of $g$ 
and $t$ for each wavelength, atmospheric conditions, and distance to the observer, provides an easy way of computing the radiance distribution in all sky directions above the observer. The angular resolution can be set with no difficulty at the level required for each application. The overall all-sky map produced by the artificial sources located in the territory surrounding the observer is built up by adding the contributions of each individual source. The contributions of sources located at a constant distance and different azimuths are just rotated versions of the same basic pattern. Note also that due to the linear nature of the atmospheric light propagation processes at the radiance levels typical of outdoor lighting systems, wide band all-sky radiance distributions in function of the distance to the observer can be computed in advance for each class of source (defined as having a given spectral and angular emission pattern) saving additional computation time.\\
The examples in the sections above are intended to provide basic insights about the performance of this approach. They correspond some simplified situations (e.g. the effects of obstacles and terrain elevation have not been included), and could be applied to more complex environments without any fundamental difficulty. Our approach is basically independent of the particular models and routines chosen by the interested researcher for computing the artificial radiance of the sky: it is  designed to provide an efficient representation of the final results, and could be applied to propagation models that would provide the optimum values of $g$ and $t$ in function of different sets of basic atmospheric and source emission parameters.




\section{Conclusions}
\label{sec:conclusions} 
We present in this work an analytical formula for the hemispherical spectral radiance produced by an artificial light source,   that allows to streamline the process of computing the all-sky artificial brightness map for observers located at any place in the world. This expression depends on two basic parameters, whose optimal values are contingent the on atmospheric conditions, source emission patterns, and distance to the observer, and can be determined by approximate analytic expressions or adequate look-up tables. This approach provides a substantial reduction in the number of free parameters required to describe the all-sky radiance produced by a ground-level elementary light source.

\section*{Acknowledgements}

This work was supported by the Slovak Research and Development Agency under contract 
No: APVV-18-0014. Computational work was supported by the Slovak National Grant Agency 
VEGA (grant No. 2/0010/20).

\section*{Data Availability Statement}
The numerical results for the all-sky radiance distributions were computed using the model developed here and that available in \citep{Kocifaj2018}. We did not use any new data.

\bsp	
\label{lastpage}
\end{document}